\documentstyle[preprint,aps]{revtex}
\tighten
\title{Nuclear Fusion via Triple Collisions in Solar Plasma }
\author{V.~B.~Belyaev, D.~E.~Monakhov, N.~Shevchenko\address{Joint
Institute for Nuclear Research, Dubna, 141980,Russia }\\
S.~A.~Sofianos, S.~A.~Rakityansky\thanks{Permanent addres: Joint
Institute for Nuclear Research, Dubna, 141980,Russia},
M.~Braun, L.~L.~Howell\address{Physics Dep., University
of South Africa, P.O. Box 392, Pretoria 0003, South Africa}\\
W. Sandhas
\address{Physikalisches Institut, Universit\"{a}t Bonn, 53115 Bonn,
Germany}}

\begin{document}
\maketitle

\begin{abstract}
 We consider several nuclear fusion reactions that take place
at the center of the sun, which are omitted in the standard
pp--chain  model. More specifically the reaction rates of
the nonradiative production of $^3$He, $^7$Be, and $^8$B nuclei
in triple collisions involving electrons are estimated within
the framework of the adiabatic approximation. These rates
are compared with those of the corresponding binary fusion reactions.
\end{abstract}
\section{}
%
There are two essential differences between  nuclear reactions
caused by binary and triple collisions. These differences can be
classified as either kinematical or dynamical.
The former are related to the selection rules which prevail in the
two-- and three--body reactions. Some binary reactions are forbidden
by conservation laws, e.g., conservation of  angular momentum, parity,
isospin etc. However such  reactions could take place in the presence
of a third particle as the three--body
mechanism is  kinematically less restricted. This may play
a significant role in  nuclear fusion in stellar plasma where the
probability of triple collisions is quite high due to
the high density of matter.

The dynamical differences stem from the interdependence
of different binary processes. For example the
${\rm e}+ {}^7{\rm Be}$ and ${\rm p}+{}^7{\rm Be}$ processes,
become dependent, when the  triple collisions
${\rm e}+{\rm p}+{}^7{\rm Be}$ in plasma are considered which leads
to a completely different physical picture.

In the present work we discuss the reaction rates
of the following processes
\begin{eqnarray}
	\rm e+p+d  & \rightarrow & \rm  {}^3He+e\ , \\
	\rm  e+p+{}^7Be &\rightarrow& \rm  {}^8B +e\ , \\
	\rm  e+{}^3He+{}^4He &\rightarrow& \rm  {}^7Be+e\ .
\end{eqnarray}
According to the standard model of the sun, the $^7$Be nucleus produced
in the pp--cycle, undergoes transmutation via two different two-body
reactions
\begin{eqnarray}
\label{eBeLin}
	{\rm e}+{}^7{\rm Be} &\to &\ {}^7{\rm Li}+\nu\ ,\\
\label{pBeBg}
	{\rm p}+{}^7{\rm Be} & \to & \ {}^8{\rm B}+\gamma\ .
\end{eqnarray}
Due to the high density of the plasma, however, electrons and protons
can always be found in the vicinity of $^7$Be and therefore
the initial  three--body quantum  state
    $ \left|{\rm p+e} + {}^7{\rm Be}\right\rangle\ $
could initiate the following reactions
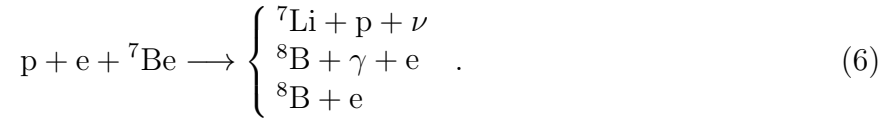
\begin{equation}
\label{3body}
	{\rm p+e} + {}^7{\rm Be}\longrightarrow\left\{
	   \begin{array}{l}
		^7{\rm Li+p}+\nu\\
		^8{\rm B}+\gamma+{\rm e}\\
		^8{\rm B+e}\\
	   \end{array}
\right.\ .
\end{equation}
In the first two reactions the proton and electron are spectators while
in the third all three initial particles are involved.
It is obvious that the three processes (\ref{3body}) are not
independent, for they all are generated by the same initial state.
As a result  the  neutrino fluxes related to $^8$B and $^7$Be nuclei are not
independent either.

The question then arises on how strong is the mutual dependence of
the processes~(\ref{3body}). This question can be formulated in
an alternative form: Up to what extent can the two--body sub-systems,
involved in (\ref{3body}), be considered as independent?
Formally such independence would imply that the
total wave function of the initial state  could be represented
by a direct product of the wave functions of the subsystems
	e+$^7$Be, p+${}^7$Be,  and  p+e,
\begin{equation}
\label{product}
	\left|{\rm p+e} + {}^7{\rm Be}\right\rangle\approx
	\left|{\rm e}+^7{\rm Be}\right\rangle\otimes
	\left|{\rm p}+^7{\rm Be}\right\rangle\otimes
	\left|{\rm p+e}\right\rangle\ .
\end{equation}
The possibility of such a factorization for three charged particles was
investigated in Refs. [1, 2], where it was shown that the total
wave function reduces to a product of the type (\ref{product}) only when all
three particles are far away from each other. Such a configuration, however,
can not contribute much to the transitions (\ref{3body}), for they are
caused by the strong and weak interactions which vanish at large distances
very fast.
Therefore the treatment of the reactions (\ref{3body})  as
independent processes requires further investigations.
\section{}
 The reaction rates of the transitions (1)-(3) can be calculated
via two approximations. In the first one  the movement
of the nuclei involved  is treated adiabatically.
This approximation should be reliable due to the mass
ratio $m_e/m_A \ll 1 $ [3].
In the second approximation  the electron--nucleus Coulomb interaction
$V_e$ could be treated perturbatively. This is based on the
fact that the corresponding Sommerfeld parameters at solar
temperatures are small. Furthermore one should bear in
mind, that in the center of the sun the average  Coulomb
interaction $<V_e>$ is of the order of 10--20\,eV which is much
smaller than the average kinetic energy ($\sim 1.5$ keV) of the
plasma particles. Therefore the higher order contributions of $V_e$
can be ignored and  the reaction rate can be written as
\begin{equation}
\label{calr}
	{\cal R}_3(\vec k , \vec k')
	=2\pi \delta( E_f-E_i)\left|\left\langle\Psi_f ,\vec k'
	\right| V_e \left|
	\Psi_i, \vec k\right\rangle
	\right|^2n_{\rm e} n_A n_B\,.
\end{equation}
Here $\vec k$ and $\vec k'$  are the electron initial
and final momenta, $n_e$ and  $n_{A,B}$ the  density of electrons
and nuclei, respectively,  while  $\Psi_i$ and $\Psi_f$ are the
 relevant initial and final nuclear wave functions.

The initial nuclear wave functions for all processes considered,
Eqs. (1)--(3), were calculated  by solving the Schr\"odinger equation using
phenomenological short range potentials and screened
Coulomb interactions. The nuclear final states for the processes (2) and (3)
were calculated  using the two-cluster models  $\rm p+^7Be $ and
 $\rm^3He + {}^4He$. The wave function of $\rm ^3 He$ was obtained
using a Faddeev-type integro-differential equation [4].

The results obtained for the ratio
\begin{equation}
\label{La}
	\Lambda_i(T)= \frac{{\cal R}^{(i)}_3(T)}{{\cal R}^{(i)}_2(T)   }
\end{equation}
for all three processes considered are presented in Table 1. In the above
${\cal R}_3(T)$ is the reaction rate (\ref{calr}) averaged over the
Maxwell distribution of the collision momenta at temperature $T$,
and ${\cal R}_2(T)$ is the corresponding value for the binary process.
The density of electrons used is  $n_e=100 N_A\;\;\mbox{\rm cm}^{-3}$,
where $N_A$ is the Avogadro number.\\[0.5cm]
Table 1.\\ {\it Ratio triple to binary reaction rates as function
of the temperature (in $10^6$ $^{\circ}$K) of the star.}
\begin{center}
\begin{tabular}{|rc|rc|rc|}
\hline
  $T_6\phantom{--}$   & $\Lambda_1(T)\times 10^3$ & $T_6\phantom{--}$ &
$\Lambda_2(T)\times 10^3$ & $T_6\phantom{--}$   & $\Lambda_3(T)\times 10^3$ \\
\hline
$1\phantom{--}  $   &  $0.218$& $14\phantom{--}$ &
$0.106$ & $10\phantom{--}$& $0.098$ \\
$5\phantom{--}  $   &  $0.147$& $15\phantom{--}$ &
$0.102$ & $12\phantom{--}$& $0.108$ \\
$10\phantom{--} $   &  $0.129$& $16\phantom{--}$ &
$0.097$ & $14\phantom{--}$& $0.139$ \\
$20\phantom{--} $   &  $0.107$& $18\phantom{--}$ &
$0.094$ & $16\phantom{--}$& $0.158$ \\
$50\phantom{--} $   &  $0.078$& $20\phantom{--}$ &
$0.089$ & $18\phantom{--}$& $0.176$ \\
$100\phantom{--}$   &  $0.055$& & & $20\phantom{--}$& $0.193$ \\
\hline
\end{tabular}
\end{center}
\vspace{0.5cm}
The binary and triple collision reaction rates increase with
increasing  temperature. However for the first two processes, (1) and
(2), the triple reaction rate grows slower as compared  to the binary one.
As a result, the ratios $\Lambda_i$ in the first two colums of
Table 1  decrease with increasing temperature.

As can be seen in Table 1 the relative contribution of the
triple processes to the fusion rates are rather small.

The temperature dependence of the binary and triple reaction rates differ
slightly. This is due to the fact that a Coulomb barrier
exists only between two particles in all the systems considered.
In contrast one can expect different  temperature dependence
in cases where all three particles are positively charged as, for example,
in the system $\rm p+p+{}^7Be$.

The two of the authors (VBB and SAR) are grateful for support received from
the Russian Foundation for Basic Research in the form of Grant No.
96 02 18678.\\

{\Large References}

[1]  G. Garibotti and J. E. Miraglia, Phys. Rev. A {\bf 21}, 572 (1980).

[2]  M. Brauner, J. S. Briggs, and H. Klar, J. Phys. B
	{\bf 22}, 2265 (1989).

[3]  S.~A.~Rakityansky, S.~A.~Sofianos, L.~L.~Howell, M.~Braun, and
     V.~B.~Belyaev\\
     \phantom{---------}Nucl.~Phys. A~{\bf 613}, 132 (1997).

[4]  M.~Fabre~de~la~Ripelle, H.~Fiedeldey, and S.~A.~Sofianos,\\
     \phantom{---------}Phys.~Rev. C~{\bf 38}, 449 (1988).
\end{document}